# NEW CONCEPT OF DYNAMIC COMPLEXITY IN QUANTUM MECHANICS AND BEYOND

**Andrei P. Kirilyuk**[*]
Institute of Metal Physics, Kiev

ABSTRACT. *The qualitatively new concept of dynamic complexity in quantum mechanics is based on a new paradigm appearing within a nonperturbational analysis of the Schrödinger equation for a generic Hamiltonian system. The unreduced analysis explicitly provides the complete, consistent solution as a set of many incompatible components ('realisations') which should permanently and probabilistically replace one another, since each of them is 'complete' in the ordinary sense. This discovery leads to the universally applicable concept of dynamic complexity and self-consistent, realistic resolution of the stagnating problems of quantum chaos, quantum measurement, indeterminacy and wave reduction. The peculiar, 'mysterious' character of quantum behaviour itself is seen now as a result of a dynamically complex, intrinsically multivalued behaviour of interacting fields at the corresponding lowest levels of the (now completely causal) structure of reality. Incorporating the results of the canonical theories as an over-simplified limiting case, this new approach urgently needs support, since its causality and completeness are directly extendible to arbitrary cases of complex behaviour of real systems, in sharp contrast to the dominating inefficient empiricism of 'computer experimentation' with primitive mechanistic (i. e. dynamically single-valued) 'models' of the irreducibly multivalued reality.*

    A fundamental, revolutionary change begins in various fields of science. It starts as a profound crisis of the canonical science, or the 'end of science' (cf. Horgan (1996)): it becomes increasingly evident that the canonical, basically linear (unitary, mechanistic) science is capable to describe only essentially one-dimensional, non-entangled (i. e. 'separable', or 'integrable'), completely predictable behaviour, whereas the irreducibly probabilistic, asymmetric, fractal-like mixture of ever developing forms constitutes the majority of the observed patterns of reality at any its level (Kirilyuk 1997).

    The canonical science tries to resolve this striking contradiction by artificial 'insertion' of the 'missing properties', i. e. by mechanistic imitation (simulation) of the natural dynamic complexity of being. This 'technique' gives rise, in particular, to the known concepts of complex behaviour and determines their unavoidable basic incompleteness and thus non-universality and the inevitable failure, even in their specific domains of origin (see e. g. Horgan (1995)). For example, dynamical chaos is interpreted as 'very rapid', but fundamentally regular (exponential) divergence of trajectories, which leads to a number of evidently inconsistent conclusions, like dependence of the resulting *fundamental* (qualitative) type of behaviour on the time of observation over the system and on other parameters. The key property of randomness, giving rise to unpredictability (or 'non-computability'), is either replaced by superficial and ambiguous 'signatures of chaos', or introduced essentially 'by hand', for example in the form of uncontrollable 'external influences', or artificial 'coarse-graining' of the degrees of freedom, which can then be 'exponentially amplified' by a chaotic system, etc. The majority of the conclusions is obtained then by a semi-empirical adjustment of such 'concepts' to the results of computer simulations performed for arbitrarily simplified abstract 'models' of the real processes.

    This fundamental impasse in the development of knowledge takes especially acute forms in quantum mechanics and the related fields dealing with the basic elements of the dynamical structure of the world. The key role of the quantum-mechanical problems is not only due to the fact that they indeed correspond to the most fundamental levels of being and therefore the essential features of quantum behaviour will inevitably be 'transmitted' to the higher, 'macroscopic' levels, but also due to the related irreducible, unambiguous character of manifestations of dynamic complexity at these lower levels, where they cannot be so easily 'simulated' by the misleading 'techniques' as it happens for the higher, less 'pure' levels. We can refer, for example, to the well-known properties of quantum unpredictability and wave-particle duality that escape any rational explanation within the canonical science and should be accepted only through postulates, simply fixing them as empirical manifestations of some fundamentally 'forbidden' (according to Niels Bohr), or effectively 'veiled' reality (see d'Espagnat (1994), (1998)).

---

[*] Address for correspondence: Post Box 115 (203/8), Kiev - 30, Ukraine 252030.
E-mail address: kiril@metfiz.freenet.kiev.ua.

Moreover, even if one accepts those contradictory postulates and concentrates on the resulting quantum dynamics as it is presented by the Schrödinger equation, the 'devil of complexity' which seemed to be well hidden within the canonical conventions inevitably reappears again at a higher level of dynamics, as the unavoidable punishment for the frustrating compromises accepted too easily. This time it takes the form of the 'problem of quantum chaos' concentrated around the truly unpredictable behaviour of a quantum system described by the formally regular, predictable Schrödinger equation and the corresponding 'unitary' (i. e. uniform) evolution.

It would be nice to discover a fundamental source of the genuine dynamical randomness in the Schrödinger equation, since then one could explain the easily observable manifestations of randomness at still higher, 'classical' levels, in accord with the well-established 'correspondence principle' proposed by Niels Bohr and eventually equivalent to the appealing statement about the intrinsic wholeness of this world, where elementary particles form atoms, atoms form larger bodies which produce still larger structures, etc. The unpleasantly striking result of the recent studies within the canonical methods unambiguously shows, however, that the conventional quantum evolution can only be unitary and thus totally predictable, reversible, etc., i. e. totally 'computable', non-complex (e. g. Ford and Mantica (1992), Ford and Ilg (1992)). Among other difficulties, this leads to the same frustrating solution of the famous Laplace paradox: once one is given a sufficiently powerful computer, one will be able to calculate every smallest motion within the universe or some its isolated part starting from its known 'initial' state, now including also the level of quantum dynamics.

The issue from this situation found by the branch of the canonical science that is properly referred to as 'quantum chaos' cannot be really considered as a 'graceful exit': whatever is its detailed version it invariably replaces the true randomness with an involved regularity or the 'unknown boundary conditions', unpredictability with the regular 'amplification' of 'veiled' initial influences, and non-computability with a quite computable, though maybe intricate, process of 'generation of random numbers'. The whole world is therefore nothing else than a huge, and very perfect generator of quasi-random numbers whose dynamics is periodic but has an 'indeed very big' period and thus practically looks random for the ephemeral human kind. Such is only one series of the far-reaching consequences of the solution to the 'problem of quantum chaos' inherent to the unitary science paradigm and known in particular as 'quantum chaology' or 'arithmetical (quantum) chaos' (e. g. Gutzwiller (1990), Giannoni, Voros, and Zinn-Justin (1991), Berry (1991), Haake (1991), Ikeda (1994), Chirikov (1995a,b), Brown (1996), Cipra (1996), Casati (1996), Zurek (1997), (1998)).

The persistence of those basically deficient solutions around quantum chaos and other problems of the omnipresent complex behaviour leads to a fundamentally 'strong' feeling that one can find a truly consistent interpretation of dynamic complexity only within some really new universal paradigm qualitatively different from the general method of the canonical science itself, but at the same time logically incorporating it as some simplified 'limiting case'. A new approach to the problem of quantum chaos generalisable to other cases of complex behaviour and possessing the above properties has been proposed recently under the reference of 'fundamental multivaluedness of dynamical functions', or 'dynamic redundance paradigm' (Kirilyuk (1995), (1996), (1997)).

The approach is demonstrated for a simple and generic example of quantum chaotic behaviour represented by a Hamiltonian system with periodic two- or three-dimensional potential of arbitrary shape and realised e. g. for a particle moving in a periodic structure or an atomic electron perturbed by the external electromagnetic wave. It is well-known that the corresponding two- or three-dimensional Schrödinger equation cannot be explicitly solved but for a very limited number of specially chosen, unrealistically symmetric potential shapes giving precisely the regular, 'exact' solutions with zero complexity, generally similar to the same kind of problem at the level of classical mechanics (the Hamilton or Newton equations). For an arbitrary shaped potential one is obliged to use a version of 'perturbation theory' based on an explicitly resolvable one-dimensional problem and considering the variation of the potential in other dimensions as 'small perturbations', whereas in reality they are not small at all, which shows the limitations of the canonical approach. Most disappointing is the fact that the obtained perturbational solutions inevitably inherit the fundamental zero-complexity deficiency of the 'exact' solutions, which largely devaluates any results within this approach, irrespective of the details.

Now in the new approach we start with the same formal constructions as the ordinary perturbative analysis, but at the critical moment when, in the ordinary approach, one is forced to dramatically simplify the situation by effectively rejecting all the dimensions but one, the new method proposes a universal way to avoid this fundamental reduction. It is possible due to an elementary algebraic transformation of the same well-known equations obtained by the canonical 'method of substitution', upon which the dynamical effects that describe essential, properly complex-dynamical behaviour can be analysed exactly, without any simplification killing complexity in the ordinary approach, whereas the unavoidable approximation is displaced towards quantities of secondary qualitative, dynamic importance.

As a result, one obtains a one-dimensional and thus formally resolvable Schrödinger equation, but with an effective potential already incorporating the specific manifestations of dynamic complexity in their unreduced quality. Namely, the effective potential - and with it the whole problem - turns out to be an intrinsically *multivalued* quantity, with its multiple individual branches generally resembling one another, but possessing at the same time considerable distinctions in the details of shape and magnitude. It is important to emphasize that each such branch of the potential and problem gives the ordinary *complete* number of solutions for the unreduced, two- or three-dimensional problem, which means that each branch provides an exhaustive image of the real system state. Therefore we call each of these complete sets of solutions, and the corresponding branch of a problem, *realisation* of the system.

It is clear that realisations are *incompatible* among them: being complete, they cannot coexist, or be 'linearly superimposed'. Since they remain at the same time *equally* real (one cannot eliminate the corresponding 'redundant' solutions by any 'legal' procedure avoiding the simplifications of perturbation theory), each of them *should* appear as a real system behaviour. The only possible issue from this apparent contradiction - indeed 'graceful' this time - provides precisely the self-consistent solution for the dynamic (quantum) chaos problem: the system should permanently and unpredictably 'switch' from one realisation to another, and the *dynamically* emerging *probability* to find the system in each of the realisations is equal to $1/N$, where $N$ is the total number of realisations that can be rigorously *obtained* as a dynamic quantity from the effective Schrödinger equation. Since the realisations can often be inhomogeneously grouped into dense agglomerates - so that experimentally the individual realisations within a group are not resolved - the probability to find the system within one of such dense groups, containing $n$ 'elementary' realisations, will be equal to $n/N$, differing therefore from the respective probabilities for other 'observable' groups.

Then *dynamic complexity* as such can be consistently defined by any strictly growing function of the number of realisations taking zero value for the case - actually exceptional - when a system has only one realisation. In this latter situation there is evidently no any transitions between realisations, and the behaviour is regular. In a typical situation, however, the system possesses more than one realisation (usually the number of realisations is much greater than one), the complexity is greater than zero and the system behaviour is *chaotic*. Therefore dynamic complexity and chaos are practically equivalent in our approach, even though there can be many qualitatively different regimes of chaos with various effective 'proportions' of regularity and randomness determined e. g. by distribution of realisations and corresponding to different observed types of complex behaviour. The whole body of qualitatively new results following from the revealed intrinsic problem multivaluedness form the concept, or paradigm of *dynamic redundance* (or *fundamental dynamic multivaluedness*).

In order to verify the validity of this our extended solution, we should test both its conceptual, logical consistence and correspondence to experimentally observed behaviour.

The conceptual soundness of the new method and its results is demonstrated by its very origin: actually we obtain the universal method of complete solution of e. g. Schrödinger equation with arbitrary potential avoiding any perturbative reductions and providing logically correct extension of the canonical, essentially one-dimensional solution to the full, intrinsically three-dimensional case. The underlying mechanism can be presented as natural, dynamically induced splitting of the system behaviour, and the corresponding adequate description, into many incompatible components. Only one of them has been taken into account before, within the canonical approach, in the form of an 'averaged' realisation, and the observed deviations from it - actually resulting from the causally random transitions to other realisations - were often ascribed to some 'noise' or indefinite 'external influences' (when they are relatively small) and sometimes referred to as 'artefacts' or (e. g. quantum) 'mysteries' (especially when they are quantitatively large and qualitatively distinct).

The 'experimental verification' of a theory of quantum chaos is not an easy task, and it will often give ambiguous results, since it is difficult to realise the corresponding pure, low-noise situation for the typically microscopic, and thus extremely sensitive, quantum systems. However, there is another universal, though indirect, way of verification based on the well-established results for the corresponding classical system (the latter can be obtained from a quantum system, for example, when the masses of the moving objects are properly increased, for the same system configuration). We actually use here the famous 'correspondence principle' already mentioned above and stating that there should exist a profoundly based quasi-continuous transition between the respective quantum and classical patterns of behaviour. This principle is evidently violated for the chaotic quantum and classical systems within the conventional concept of quantum chaos, which demonstrates its weakness (even though the canonical concept of chaos at classical levels shows itself, upon closer examination, a similar inconsistency of fundamental origin). In our approach this basic contradiction of the conventional 'chaology' is not only completely removed, but is actually transformed into a convincing evidence in favour of the multivalued solution. Indeed, our purely quantum-mechanical, analytical description reproduces the characteristic classically established regimes of global chaos, regularity and the transition between the two with changing parameters, including the expression for the point of transition and the known regime of asymptotically weak chaos in the 'stochastic' layer within the regime of global regularity.

Another possible way of practical verification of a quantum chaos description consists in modelling of a quantum system by electromagnetic waves in a reflecting cavity. Such experiments have been performed, but they were oriented to the canonical theory operating with abstract 'signatures' of chaos like certain statistical laws of energy level distribution, etc. Sometimes these experiments correspond to the ground level of a quantum system, which is just the exceptional case where, according to our description, the true quantum chaos (realisation change) is absent, since the number of active realisations is reduced to one and the complexity is zero (this explains, by the way, the practically important stability of the ordinary matter, composed from many interacting particles, in its ground state). Therefore, in order to discover the explicit signs of the true 'quantum' chaos in the behaviour of electromagnetic waves in resonators (and actually in any other system) one should be capable to observe the detailed, non-stationary behaviour of a system with more than one effective dimension excited over its ground state (even though in more sophisticated, many-body systems the ground state can also be intrinsically chaotic).

Finally, direct and unambiguous experimental manifestations of the true quantum chaos can probably be observed in a particular and practically important case of particle channeling in crystals, as it was shown in an earlier paper (Kirilyuk 1992).

It is not difficult to see from the underlying analysis that the proposed new concept of quantum chaos and dynamic complexity in quantum mechanics can be directly generalised to arbitrary complex behaviour of any kind of system (Kirilyuk 1997).

Thus, application of the same approach to quantum systems at somewhat more fundamental level of dynamics than that of the Schrödinger equation provides a consistent, causal solution to the problems of 'quantum measurement' explaining the famous 'quantum mystery' of the wave-particle duality, involving 'wave collapse' and 'quantum indeterminacy', as a particular case of complex behaviour, physically and mathematically transparent (Kirilyuk 1995, 1997). Quantum measurement is interpreted now as any nontrivial interaction of elementary *physical* waves which leads to their *real* collapse (a specific self-sustained dynamical squeeze) towards a 'compact' state representing here a particular system realisation and localised around a dynamically unpredictable 'centre of reduction', while the above naturally deduced expression for probabilities corresponds simply to the well-known 'Born's probability rule', actually postulated in the conventional quantum mechanics. Therefore the basic transformation of a quantum wave into particle and back, together with its 'quantum indeterminacy', discussed so passionately by Bohr, Einstein and other founders of quantum mechanics, is consistently explained now by the above dynamic uncertainty of multiple incompatible (redundant) realisations represented here by different possible centres of wave collapse and the extended wave itself. The latter can therefore be considered now as a real, physical wave obeying the causally *derived* Schrödinger equation, which puts an end to the long-standing ambiguity around quantum wave, particle, and their duality. The basic notion of 'quantum' or 'quantization' (a specific dynamic discreteness of the micro-object behaviour) is causally understood now as an elementary manifestation of the universal dynamic complexity in the form of dynamical splitting into *discrete* realisations and permanent transitions between them. This kind of behaviour emerges starting from the most fundamental level of the perceivable world, where the simple attractive interaction between two physically real media, the electromagnetic and gravitational 'protofields', provides the dynamically chaotic internal dynamics of the elementary fermion (electron) with the universally interpreted property of (relativistic) mass-energy accounting for both inertia and universal gravitation (Kirilyuk 1997). This is the direct extension of the unreduced version of the 'double solution' proposed by Louis de Broglie (see de Broglie (1956), (1964), (1971), (1976)) and incorporating the unified and now mathematically correct versions of both physically real wave-particle duality (dynamic redundance!) and 'hidden thermodynamics of a particle' (intrinsic dynamical chaos!).

Ascending to higher levels of thus naturally emerging and interacting objects-realisations and applying always the same description of dynamically complex, redundantly multivalued behaviour, we obtain the 'universal hierarchy of complexity' opening a way to the generalised, holistic description of the world dynamics within the single unifying concept (Kirilyuk 1997). The obtained world image reproduces the properties of reality we see around us in their full intricacy and diversity, in sharp contrast to the mechanistic approach of the canonical, linear (= single-valued) science. The World in the whole, and any its particular part and level, including the cases of quantum chaos and quantum measurement, are *intrinsically unpredictable* ('non-computable') now, and any regular, ordinary computation cannot autonomously reproduce even most simple cases of chaotic behaviour (like that of a quantum or classical particle in a realistically shaped potential well), whatever is its formal power. This is the definite, well specified end of the long dominance of the mechanistic, dully repetitive approach in science and eventually the end of the medieval, reductive way of thinking in general, far beyond the usual interpretation of 'science'.

# References

Berry, M.V. (1991). In *Chaos and Quantum Physics*, ed. M.-J. Giannoni, A. Voros, and J. Zinn-Justin (North-Holland, Amsterdam).

de Broglie, L. (1956). *Une Tentative d'Interprétation Causale et Non-Linéaire de la Mécanique Ondulatoire* (Gauthier-Villars, Paris).

de Broglie, L. (1964). *La Thermodynamique de la Particule Isolée (Thermodynamique Cachée des Particules)* (Gauthier-Villars, Paris).

de Broglie, L. (1971). *La Réinterprétation de la Mécanique Ondulatoire.* 1re partie: *Principes Généraux* (Gauthier-Villars, Paris).

de Broglie, L. (1976). *Recherches d'un Demi-Siècle* (Albin Michel, Paris).

Brown, J. (1996). "Where two worlds meet...", New Scientist **150**, No. 2030 (18 May), p. 26.

Casati, G. (1996). Chaos **6**, 391.

Chirikov, B.V. (1995a). "Pseudochaos in Statistical Physics", Preprint of the Budker Institute of Nuclear Physics, Budker INP 95-99 (Novosibirsk). Also: e-print chao-dyn/9705004.

Chirikov, B.V. (1995b). "Linear and Nonlinear Dynamical Chaos", Preprint of the Budker INP 95-100 (Novosibirsk). Also: e-print chao-dyn/9705003.

Cipra, B. (1996). "Prime formula weds number theory and quantum physics", Science **274**, 2014.

d'Espagnat, B. (1994). *Le Réel Voilé* (Fayard, Paris). English version: *Veiled Reality. An Analysis of Present-Day Quantum Mechanical Concepts* (Addison-Wesley, Reading, 1995).

d'Espagnat, B. (1998). "Quantum Theory: A Pointer to an Independent Reality", e-print quant-ph/9802046.

Ford, J. and M. Ilg (1992). Phys. Rev. A **45**, 6165.

Ford, J. and G. Mantica (1992). Am. J. Phys. **60**, 1086.

Giannoni, M.-J., A. Voros, and J. Zinn-Justin (eds.) (1991). *Chaos and Quantum Physics* (North-Holland, Amsterdam).

Gutzwiller, M.C. (1990). *Chaos in Classical and Quantum Mechanics* (Springer-Verlag, New York).

Haake, F. (1991). *Quantum Signatures of Chaos* (Springer-Verlag, Berlin).

Horgan, J. (1995). "From Complexity to Perplexity", Scientific American, June 1995, 74.

Horgan, J. (1996). *The End of Science* (Addison-Wesley, Helix).

Ikeda, K. (ed.) (1994). *Quantum and Chaos: How Incompatible ?* Proceedings of the 5th Yukawa International Seminar, Progr. Theor. Phys. Suppl. No. 116.

Kirilyuk, A.P. (1992). Nuclear Instrum. Meth. **B69**, 200.

Kirilyuk, A.P. (1995). "Causal Wave Mechanics and the Advent of Complexity", Parts I-V, e-print quant-ph/9511034-38. Also: "Quantum Mechanics with Chaos: Correspondence Principle, Measurement and Complexity" (Preprint of the Institute of Metal Physics No. 95-1, Kiev, 1995), 116 p.; e-print chao-dyn/9510013.

Kirilyuk, A.P. (1996). Annales de la Fondation L. de Broglie **21**, 455.

Kirilyuk, A.P. (1997). *Universal Concept of Complexity by the Dynamic Redundance Paradigm: Causal Randomness, Complete Wave Mechanics, and the Ultimate Unification of Knowledge* (Naukova Dumka, Kiev).

Zurek, W.H. (1997). Phys. World **10**, No. 1 (1997) 24.

Zurek, W.H. (1998). "Decoherence, Chaos, Quantum-Classical Correspondence, and the Algorithmic Arrow of Time", e-print quant-ph/9802054.